\documentclass[prb,twocolumn,showpacs,floatfix,preprintnumbers,amsmath,amssymb,superscriptaddress]{revtex4}

\usepackage{graphicx}
\usepackage{amsmath}
\usepackage{amssymb}
\usepackage{color}
\usepackage{dcolumn}
\usepackage{epsfig}
\usepackage{bm}
\usepackage{subfigure}
\usepackage[urlcolor=blue]{hyperref}
\hypersetup{backref, colorlinks=true, linkcolor=blue, citecolor=blue}

\begin{document}

\title{Thermoelectric properties of polycrystalline palladium sulfide}

\author{Liu-Cheng Chen}
\affiliation{Institute of Solid State Physics, Chinese Academy of Sciences, Hefei 230000, China}
\affiliation{University of Science and Technology of China, Hefei 230026, China}
\affiliation{Center for High Pressure Science and Technology Advanced Research, Shanghai 201203, China}

\author{Bin-Bin Jiang}
\affiliation{State Key Laboratory of High Performance Ceramics and Superfine Microstructure, Shanghai Institute of Ceramics, Chinese Academy of Science, Shanghai 200050, China}
\affiliation{University of Chinese Academy of Sciences, Beijing 100049, China}

\author{Hao Yu}
\affiliation{Center for High Pressure Science and Technology Advanced Research, Shanghai 201203, China}

\author{Hong-Jie Pang}
\affiliation{Center for High Pressure Science and Technology Advanced Research, Shanghai 201203, China}

\author{Lei Su}
\affiliation{Key Laboratory of Photochemistry, Institute of Chemistry, Chinese Academy of Sciences, Beijing, 100080, China}

\author{Xun Shi}
\affiliation{State Key Laboratory of High Performance Ceramics and Superfine Microstructure, Shanghai Institute of Ceramics, Chinese Academy of Science, Shanghai 200050, China}

\author{Li-Dong Chen}
\affiliation{State Key Laboratory of High Performance Ceramics and Superfine Microstructure, Shanghai Institute of Ceramics, Chinese Academy of Science, Shanghai 200050, China}

\author{Xiao-Jia Chen}
\email{xjchen@hpstar.ac.cn}
\affiliation{Center for High Pressure Science and Technology Advanced Research, Shanghai 201203, China}

\date{\today}

\begin{abstract}
A suite measurements of the electrical, thermal, and vibrational properties are conducted on palladium sulfide (PdS) in order to investigate its thermoelectric performance. The tetragonal structure with the space group $P$42/$m$ for PdS is determined from X-ray diffraction measurement. The unique temperature dependence of mobility suggests that acoustic phonons and ion impurity scattering are two dominant scattering mechanisms within the compound. The obtained power factor of $27$ $\mu$Wcm$^{-1}$K$^{-2}$ at 800 K is the largest value in the remaining transition-metal sulfides studied so far. The maximum value of the dimensionless figure of merit is 0.33 at 800 K. The observed phonon softening with temperature indicates that the reduction of the lattice thermal conductivity is mainly controlled by the enhanced lattice anharmonicity. These results indicate that the binary bulk PdS has promising potential to have good thermoelectrical performance.
\end{abstract}


\maketitle

\section{INTRODUCTION}

Thermoelectric materials, which can directly convert heat to electrical power, have been of interest for many years, because of their potential applications with environment-friendly properties. The efficiency of thermoelectric materials is determined by the dimensionless figure of merit ($zT$), defined as $zT$ = $\frac{S^2\sigma}{\kappa}T$, where $S$ is the Seebeck coefficient, $\sigma$ is the electrical conductivity, $T$ is the absolute temperature, and $\kappa$ is the thermal conductivity. For thermoelectric devices, a conversion efficiency of 15\% ($zT\geq$1) is needed through obtaining large $S$, high $\sigma$, and low $\kappa$\cite{bell,snyder}. Many strategies for enhancing $zT$ have been proven to be effective, including phonon-liquid electron-crystal, nanostructure engineering, and band structure engineering\cite{sales,biswas,ypei}. Among them, reduction of the thermal conductivity by micro-structure modification is the ripest strategy. Therefore, searching for thermoelectric materials with intrinsic large power factor $PF$($PF=S^2\sigma$), and then modified through nanostructuring while maintaining
the original electrical properties will be a good method for superior performance of thermoelectric materials.

Recently, transition-metal sulfides have attracted much attention, because of the cheaper and earth abundant element of sulfide compared with telluride or selenium\cite{luxu,wancl}. Many transition-metal sulfides have shown good thermoelectric performance, mainly because of their relatively low thermal conductivities. For instance, the highest $zT$ of PbS has been improved from 0.4 to $\sim$1 due to the low thermal conductivity through micro-structure modification\cite{zhaold2,johnsen1,zhaold}. The reduction of thermal conductivity leads to a high $zT$ of 0.6 at 873 K for SnS through the method of doping\cite{qingt}. Copper sulfide is an important thermoelectric material with high $zTs$ ($zTs=1.4-1.7$ at 1000 K). The ultralow lattice thermal conductivities caused by the liquid-like copper ions were proposed to account for such high $zTs$ in Cu$_x$S\cite{heying}. However, the $PFs$ of those thermoelectric sulfides are not high for providing the ideal $zTs$. For example, the best performance material Cu$_{1.97}$S only has a $PF$ near 8 $\mu$Wcm$^{-1}$K$^{-2}$ at 1000 K. Thus, an interesting approach to give a high $zT$ value is to search a thermoelectric sulfide with intrinsic large $PF$. Palladium sulfide (PdS), which belongs to transition-metral sulfide, has potential applications in semiconducting, photoelectrochemical, and photovoltaic fields, because of its ideal band gap of 1.6 eV\cite{Folmer,Ferrer,Barawi}. Furthermore, it also has several potential device applications in catalysis and acid resistant high-temperature electrodes\cite{Bladon,Zubkov,yang1}. In a study of the thermoelectric properties\cite{Pascual}, the Seebeck coefficient of PdS thin film was reported to have a large value (about 280 $\mu$V/K) at room temperature. Therefore, in this experimental campaign, it is highly desired to investigate its bulk thermoelectric properties with the purpose of exploring the viability of the sample as potentially useful thermoelectric material. In addition, it is interesting to gain some insight into the physical mechanisms for enhancing $zT$ of thermoelectric materials.

In this work, PdS polycrstal was successfully fabricated by melting-quenched and spark plasma sintering technique. The thermoelectric properties were investigated with the measurements of electrical conductivity, Seebeck coefficient, thermal conductivity. We found that the power factor has a very large value ($PF$=27 $\mu$Wcm$^{-1}$K$^{-2}$) derived from high $\sigma$ and large $S$. The highest $zT$ value achieved in this paper is 0.33 at 800 K. Our results highlight that the polycrystalline PdS is a promising potential thermoelectric material through overcoming the high $\kappa$ with micro-structure modification.

\section{EXPERIMENTAL DETAILS}

High purity raw elements, Pd (powders, 99.99\%, Alfa Aesar) and S (powders, 99.999\%, Alfa Aesar) were weighted out in the stoichiometric proportions and then mixed well in an agate mortar. The mixtures were pressed into pellets and sealed in quartz tubes under vacuum. Then the tubes were heated at a speed of 1 $^{\circ}$C/min to 1373 K and remained at this temperature for 12 hours before quenched into cold water. Next, the quenched tubes were annealed at 873 K for 7 days. Finally, the products were ground into fine powders and sintered by Spark Plasma Sintering (Sumitomo SPS-2040) at 923 K under a pressure of 65 Mpa for 5 mins. High-density samples ($>$99\% of the theoretical density) were obtained.

\begin{figure}[tbp]
\includegraphics[width=\columnwidth]{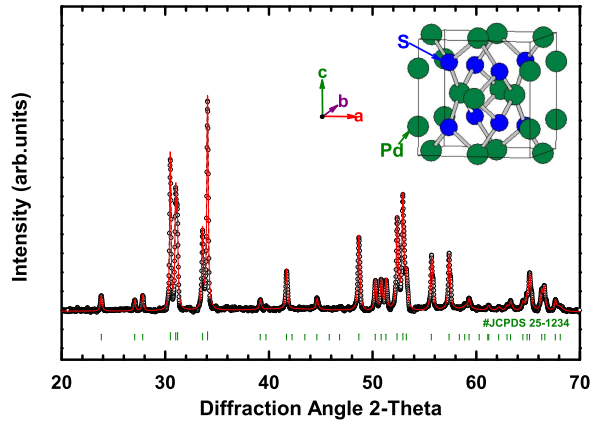}
\caption{Power XRD pattern of the PdS sample at room temperature. The inset illustrates the crystal structure of PdS.}
\end{figure}

The powders were characterized by X-ray diffraction (XRD) (Rigaku, Rint 2000) with Cu $K\alpha$ radiation ($\lambda$ = 1.5405 \r{A}) at room temperature. The measurements were performed between 20$^{\circ}$ and 70$^{\circ}$ with a scan width of 0.02$^{\circ}$ and rate of 2$^{\circ}$/min. Seebeck coefficient, electrical and thermal conductivity, were simultaneously measured between 3 K and 300 K in a thermal transport option (TTO) setup within a Physical Properties Measurements System (PPMS) by Quantum Design. The measurements were carried out in the residual vacuum of a He atmosphere, under a pressure of 10$^{-5}$ Torr. The typical size of PdS used in PPMS was 4.3$\times$2.0$\times$0.9 mm$^3$ with four Cu wires attached with Ag paste. Hall coefficient, R$_H$, was also measured with conventional four-probe technique by PPMS with a temperature range of 7 K to 300 K. Heat capacity, C$_p$, in the temperature range of 1.8 to 300 K was additionally measured by PPMS in order to analyze the thermal conductivity data. To complete the measurements, Raman scattering spectra of PdS were obtained by a system with using a 488 nm laser with a stabilized power of 3 mW. The Charge Coupled Device was manufactured by Princeton Instruments. The low-temperature measurements of the Raman setup were performed in a range of 10 K and room temperature, using a liquid helium cryostat. For the high temperature Raman spectra, a graphite flake was used for heating in Diamond Anvil Cell and the temperature was controlled by a K-type thermocouple. The high temperature electrical conductivity and the Seebeck coefficient were measured by using an Ulvac ZEM-3 from 300 to 800 K under a Helium atmosphere. The high temperature thermal conductivity was calculated from $\kappa$ = $DC_p$$\rho$, where the thermal diffusivity ($D$) was obtained by using a laser flash method (Netzsch LFA 457) under an Argon atmosphere. The specific heat ($C_p$) was calculated by using Dulong-Petit law, and the density ($\rho$) was measured using the Archimedes method.

\section{RESULTS AND DISCUSSION}

\subsection{Electrical transport properties}

\begin{figure*}[tbp]
\includegraphics[width=\textwidth]{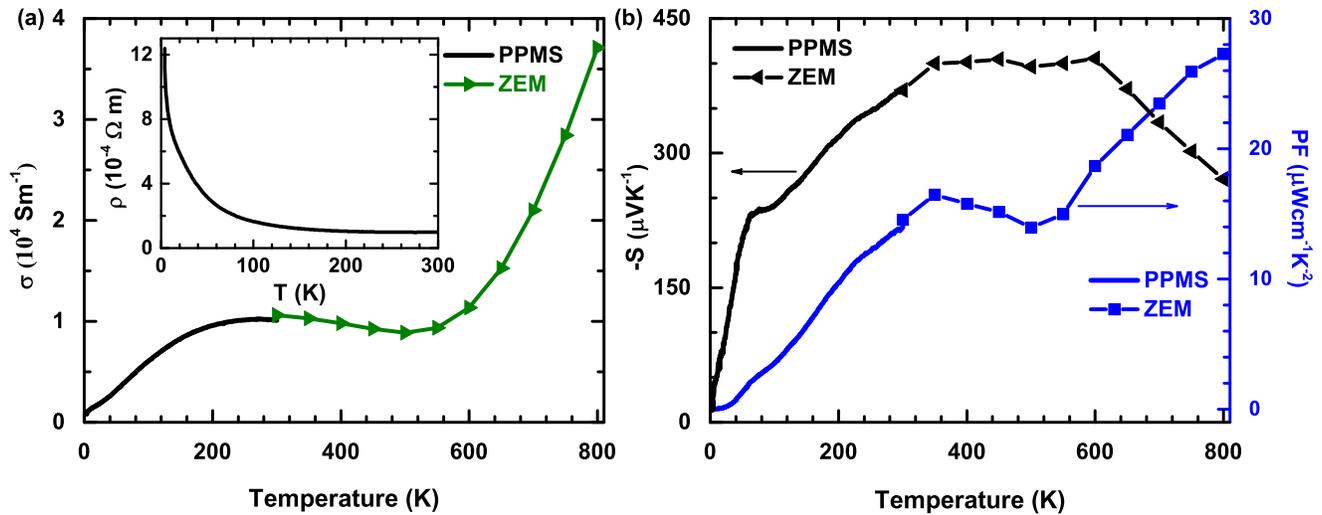}
\caption{(color online). Temperature dependence of the electrical transport properties of bulk PdS; (a) Temperature dependence of the electrical conductivity $\sigma$ from 3 to 800 K. Inset shows the $\rho$ versus $T$ below room temperature. (b) Temperature dependence of the Seebeck coefficient ($S$) and the calculated power factor (PF = $S^2\sigma$) from 3 to 800 K.}
\end{figure*}

The XRD pattern and crystal structure of bulk PdS are shown in Fig. 1. The main diffraction peaks indicated a tetragonal structure (JCPDS No.25-1234) with the space group $P$42/$m$ (84), no other phase being detected obviously from the XRD patterns. The lattice parameters are $a$=$b$=6.016 \AA, $c$=6.63 \AA.

Temperature dependence of $\sigma$, $S$, and $PF$ of polycrystalline PdS is shown in Fig. 2. It can be seen that whether the $\sigma$, $S$, or $PF$ measured in the different studies are consistent with each other. The small deviations of data around around the room temperature could be caused by the different error of the two systems. In Fig. 2(a), the behavior of $\sigma$ with temperature is complex. At high temperatures, $\sigma$ decreases sharply with decreasing temperature until around 500 K, which is probably caused by the thermal activation induced bipolar effect in semiconductors. However, in the temperature range of 200 - 500 K, $\sigma$ behaves likely a constant. It does not show an obvious increase with decreasing temperature. This behavior is consistent with that of a metal, which was not observed previously\cite{Ferrer}. Below 200 K, there is a significant decrease in the temperature dependence of $\sigma$, decreasing sharply as the sample is cooled. This reflects the nature of a typical semiconductor. These behaviors at low temperatures can be seen more obviously from the inset [Fig. 2(a)]. Therefore, these contrasting behaviors imply that the bulk PdS may have semiconductor-metal and metal-semiconductor transitions around 200 K and 500 K, respectively.

In Fig. 2(b), the temperature dependences of $S$ and $PF$ are shown, respectively. The value of $S$ is negative, indicating that the majority of the charge carries are electrons. With increasing temperature, the absolute value of $S$ increases to a vary large value approximately 400 $\mu$VK$^{-1}$ without doping at about 350 K, and then the tendency behaves like a platform until about 600 K. Unfortunately, it has an obvious decrease above 600 K, which may be caused by thermal excitation of carriers. The phenomenon of $\sigma$ and $S$ at high temperatures has been observed in many other intrinsic semiconductor systems which may be ameliorated by doping\cite{YPei}. Notably, the value of $PF$ has an increasing tendency over the entire temperature range, except the intermediate region. The maximum value of $PF$ is about 27 $\mu$Wcm$^{-1}$K$^{-2}$ at 800 K (the highest recorded temperature in this study). Here, the value of $PF$ is very large compared with other thermoelectric sulfides, such as about three times larger than the value of optimal Cu$_{1.97}$S\cite{heying}. From the electrical properties, it is obvious that PdS is a potentially useful thermoelectric material.

Hall effect was measured which could give an insight into the electrical transport properties. The temperature dependence of the carrier concentration ($n_H$) and Hall mobility ($\mu_H$) evaluated from low temperature Hall measurements are shown in Fig. 3. The Hall coefficient (R$_H$) is negative over the entire temperature range, indicating the majority of the charge carries are electrons, which is consistent with the negative Seebeck coefficient. The temperature dependent $\mu_H$ has a gentle increasing until around 200 K, and the maximum value is 230 cm$^2$V$^{-1}$s$^{-1}$. The relatively high value of $\mu_H$ could be caused by the covalent bond characteristics of the tetragonal structure. Then it keeps almost unchanged up to 300 K. This behavior is consistent with the temperature dependence of $\sigma$ below 300 K as shown in Fig. 2(a). The scattering mechanisms that dominate the transport properties in bulk PdS below 300 K are long-wavelength acoustic phonons and ion impurity scattering, indicated by fitting $\mu_H$ using the rough relation\cite{Ravich} $\mu_H \propto AT^{-2/3} + BT^{2/3}$. The temperature dependent $n_H$ is very complex and incredible, especially at low temperatures, which is most likely caused by some magnetic transitions. However, the carrier concentration of PdS changes in the same order, which indicates that the carrier concentration for PdS has a weak temperature dependence below 300 K, which has also been observed in lead chalcogenides\cite{Scanlon}.

\begin{table*}[tbp]
\caption{Summary of the thermal diffusivity $D$, heat capacity $C_p$, and density $d$, which were used to calculate the thermal conductivity $\kappa$ of PdS.}
\begin{tabular}{c|ccccccccccc}
\hline\hline
\textbf{$T$ (K)} & 300 & 350 & 400 & 450 & 500 & 550 & 600 & 650 & 700 & 750 & 800\\
\hline
\textbf{$D$ (cm$^2$/sec)} & 10.137 & 8.328 & 7.046 & 6.015 & 5.216 & 4.602 & 4.083 & 3.67 & 3.305 & 3.035 & 2.817\\
\hline
\textbf{$C_p$ (Jg$^{-1}$K$^{-1}$)} & 0.36 & 0.36 & 0.36 & 0.36 & 0.36 & 0.36 & 0.36 & 0.36 & 0.36 & 0.36 & 0.36\\
\hline
\textbf{$d$ (g/cm$^3$)} & 6.6 & 6.6 & 6.6 & 6.6 & 6.6 & 6.6 & 6.6 & 6.6 & 6.6 & 6.6 & 6.6\\
\hline\hline
\end{tabular}
\end{table*}

\subsection{Heat transport properties}

\begin{figure}[b]
\includegraphics[width=\columnwidth]{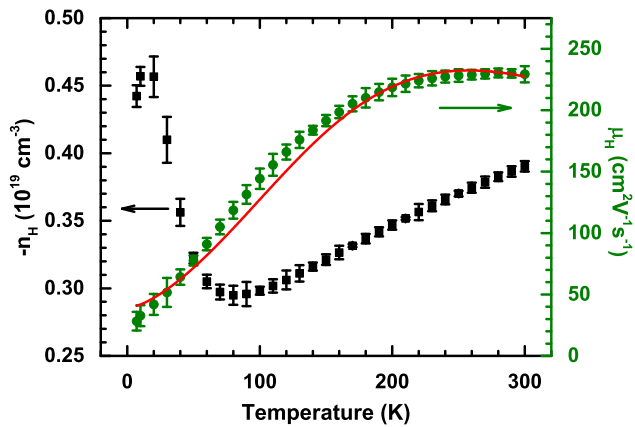}
\caption{Temperature dependence of the Hall carrier concentration ($n_H$) for PdS from 6 K to 300 K (black line) and the Hall mobility ($\mu_H$) (olive line) at the same temperature range, The red solid line is the fitted curve for $\mu_H \propto AT^{-2/3} + BT^{2/3}$.}
\end{figure}

The temperature dependence of the thermal diffusivity and specific heat capacity used for the $\kappa$ calculation are shown in Table I. The obtained $\kappa$ with temperature is shown in Fig. 4. It can be seen that $\kappa$ increases sharply with increasing temperature and evolves through a maximum (about 130 Wm$^{-1}$K$^{-1}$) at 38 K before finally decreasing roughly in a T$^{-1}$ relation. This phenomenon indicates that the bulk PdS is a normal crystal compound. Generally, $\kappa$ consists of the electronic part $\kappa_{ele}$ and the lattice part $\kappa_{lat}$. The electronic part $\kappa_{ele}$ is proportional to the electrical conductivity $\sigma$ through the Wiedemann-Franz relation\cite{Kumar}: $\kappa_{ele}$=$L\sigma T$, where $L$ is the Lorenz number ($L$=2.44$\times$10$^{-8}$W$\Omega$K$^{-2}$ in theory for semiconductor). Here, $\kappa_{ele}$ could be ignored below room temperature as shown in the inset of Fig. 4. However, the contribution of $\kappa_{ele}$ to the total $\kappa$ increases with increasing temperature and reaches about 12\% at 800 K. The high $\kappa$ of PdS probably comes from the light atomic mass of sulfur and the strong chemical bonds in the crystal. In order to elucidate the reasons for the high thermal conductivity of PdS, heat capacity $C_p$ at low temperatures was measured. The results are shown in Fig. 5. The measured $C_p$ value at 300 K is 0.35 Jg$^{-1}$K$^{-1}$, which is closed to the theoretical value (0.36 Jg$^{-1}$K$^{-1}$). The inset of Fig. 5 displays the heat capacity $C_p$ $vs.$ $T$ representation. The solid red line is the fitted curve based on the Debye mode by the relation\cite{Gofryk}: $C_p$ = $\varphi$$T$ + $\beta$$T^3$. The total $C_p$ includes the carrier contribution, $\varphi$$T$, and the phonon contribution, $\beta$$T^3$. The fitted parameters are 0.34 mJmol$^{-1}$K$^{-2}$ for $\varphi$ and 0.13 mJmol$^{-1}$K$^{-2}$ for $\beta$. The small $\varphi$ indicates the electronic density of states near the Fermi level is quite weak at low temperatures when compared with other thermoelectric materials ($e.g.$ YbFe$_4$Sb$_{12}$, $\varphi$ = 141.2 mJmol$^{-1}$K$^{-2}$)\cite{Dilley}. This finding is consistent with the low $\sigma$ observed at low temperatures demonstrated in Fig. 2(a).

\begin{figure}[b]
\includegraphics[width=\columnwidth]{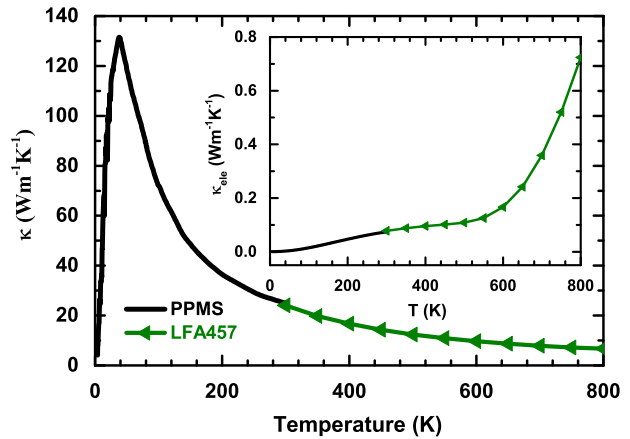}
\caption{Temperature dependence of the total thermal conductivity $\kappa$ of bulk PdS. The inset illustrates the electric thermal conductivity $(\kappa_{ele})$ $vs.$ $T$ plot in the temperature range between 3 and 800 K.}
\end{figure}

It is known that the Gr\"{u}neisen parameter ($\gamma$) reflects the strength of the lattice anharmonicity, and generally a large $\gamma$ value coupled with a strong phonon-phonon interaction results in the low lattice thermal conductivity. The Gr\"{u}neisen parameter can be fitted according to the formula: $\omega = \omega_0(1 + 3\alpha)^{-\gamma}$, ($\omega_0$ is phonon frequency at 0 K and $\alpha$ is thermal expansion coefficient)\cite{DHwang}. Therefore, the temperature-dependent spectra of Raman scattering were measured. Figure 6 shows the frequency change of the obtained Raman modes as a function of temperature. To clarity, we denote those Raman modes as L$_n$, which should correspond respectively to the lattice modes of bulk PdS. All of the lattice modes exhibit softening behavior from 10 to 523 K, which suggests that the lattice anharmonicity is gradually stronger with increasing temperature\cite{cwli}. This phenomenon is mainly responsible for the decreasing $\kappa$ at high temperatures.

An obvious transition can be seen around 350 K from Raman spectra which was labeled by the shadow region as shown in Fig. 6(b). Accordingly, the temperature dependence of the obtained phonon frequencies was divided into two segments for fitting in order to obtain more reasonable Gr\"{u}neisen parameters. The results of the fitted $\gamma_1$ and $\gamma_2$ for detected phonon modes are listed in Table II. The results give new insights into the understanding of high $k_{lat}$ observed in this system. The fitted values of $\gamma_1$ of the low frequency modes $L_1$ $\sim$ $L_6$ have relatively larger Gr\"{u}neisen parameters, which means that the six mode phonons are more sensitive to be scattered with temperature changing compared with the other lattice modes in the temperature range of 10 $\sim$ 350 K. The smaller values of $\gamma_1$ compared with $\gamma_2$ correspond to the high $\kappa_{lat}$ at low temperatures. Surprisingly, the lattice mode L$_9$ disappears at 200 K, which is not reflected in the thermal conductivity at the same temperature as shown in Fig. 4. This case illustrates that the lattice mode L$_9$ has a little influence to the thermal conductivity. Between 350 and 523 K, the fitted Gr\"{u}neisen parameters ($\gamma_2$) at high temperatures have a visible difference compared with $\gamma_1$ at low temperatures.

From Table II, it is obvious that the Gr\"{u}neisen parameters of $L_1$ $\sim$ $L_4$ above 350 K become weaker, however, the other phonon modes become stronger, especially, for the phonon mode $L_6$ ($\gamma_2$ = 1.90). This illustrates that the strength of the lattice anharmonicity becomes stronger at high temperatures. This behavior is consistent with the dramatical decrease of $\kappa_{lat}$ in the same temperature range. However, the Gr\"{u}neisen parameters are still very small even at high temperatures compared with typical thermoelectric materials, such as SnSe ($\gamma$=4.1 in $a$ axes)\cite{dongz}, and AgSbSe$_2$ ($\gamma$=3.5)\cite{Nielsen}. Thus, it is reasonable that increasing phonon scatting which contributes to the complex phonon density of states in crystal may be an effective way to reduce lattice thermal conductivity.

\begin{figure}[tbp]
\includegraphics[width=\columnwidth]{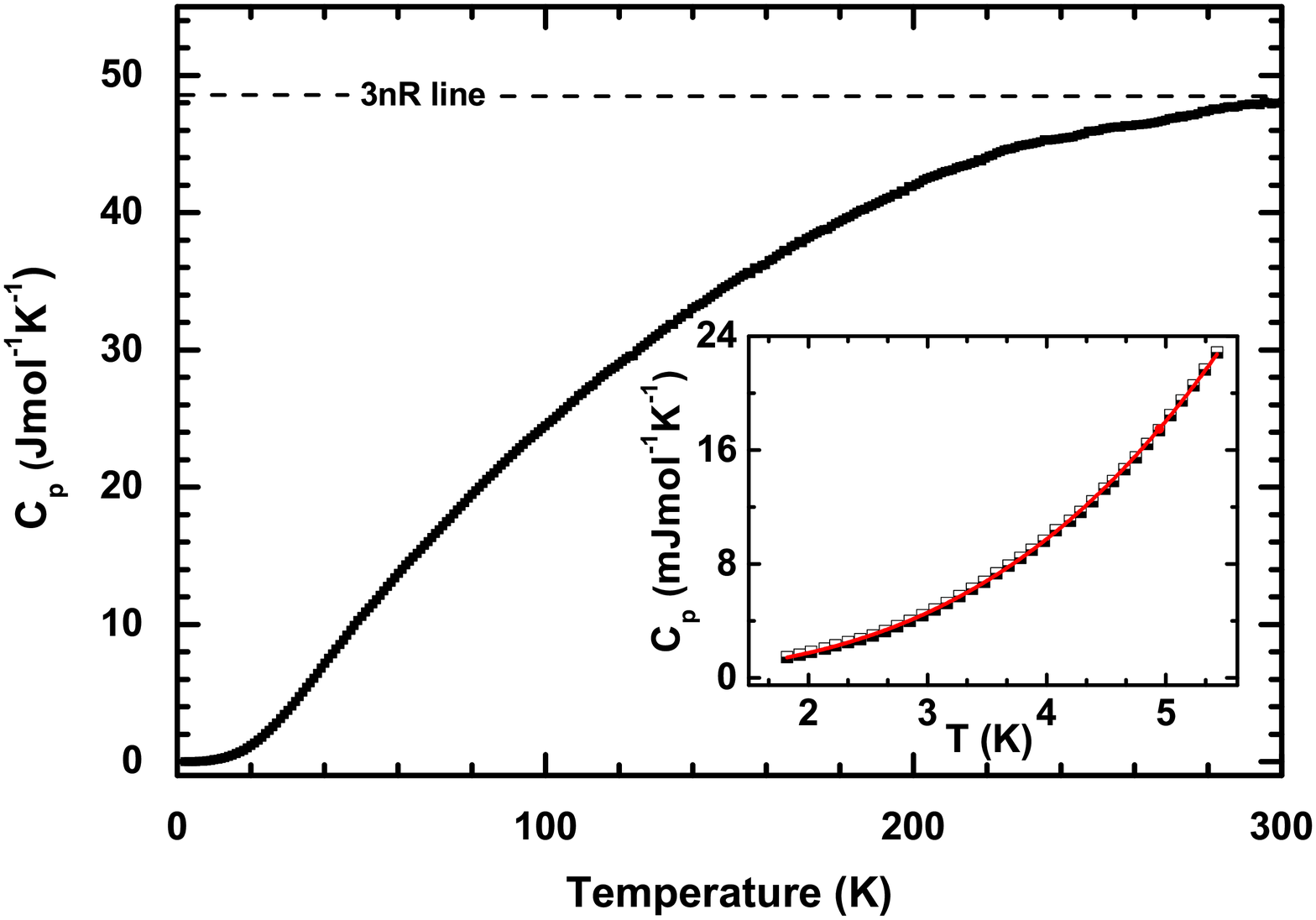}
\caption{Temperature dependence of the heat capacity ($C_p$) for bulk PdS from 1.8 K to 300 K. The dash line is the 3nR line calculated by the Dulong-Petit model. The inset shows the heat capacity below 6 K. The solid red line is the fitted curve by using $C_p$ = $\varphi$$T$ + $\beta$$T^3$.}
\end{figure}

\subsection{The dimensionless figure of merit}

Based on the measured temperature-dependent $S$, $\sigma$, and $\kappa$, the dimensionless figure of merit $zT$, which directly determines the energy conversion efficiency of a thermoelectric material, is calculated. The results are shown in Fig. 7. The calculated value of $zT$ has a positive temperature dependence over the entire measurement range, different from other thermoelectric materials which have a maximum peak at some suitable temperature. This phenomenon indicates that the thermoelectric properties of bulk PdS will be more efficient at higher temperature. The value of $zT$ reaches to 0.33 at temperature of 800 K (the highest recorded temperature in this study). It is considerably high compared with many undoped thermoelectric sulfides at around the the similar temperatures, such as Bi$_2$S$_3$ ($zT$ $\sim$ 0.2 at 823 K)\cite{zhge}. However, the thermoelectric conversion efficiency is still low compared with high performance thermoelectric materials, even it has an obvious uptrend at higher temperatures. Thus, it is extremely urgent to improve the thermoelectric conversion efficiency for bulk PdS.

\begin{figure}[tbp]
\includegraphics[width=\columnwidth]{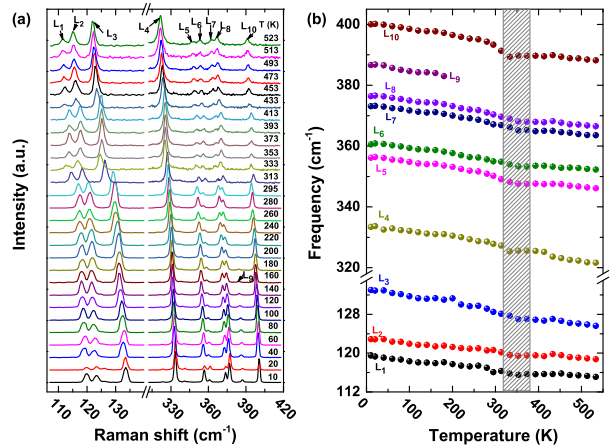}
\caption{(a) Raman scattering spectra of bulk PdS at the temperature range of 10 K and 523 K. (b) Temperature dependence of the obtained phonon modes.}
\end{figure}

\begin{table}[b]
\caption{The frequencies (cm$^{-1}$) of observed Raman modes and the fitted Gr\"{u}neisen parameters ($\gamma_1$, $\gamma_2$) of bulk PdS.}
\begin{tabular}{c|c|c|c}
\hline\hline
\textbf{Raman modes} & \textbf{Obs (cm$^{-1}$)} & \textbf{$\gamma_1$} & \textbf{$\gamma_2$} \\
\textbf{L$_n$} & \textbf{at 10 K} &  & \\
\hline
  L$_{1}$ & 119.3 & 0.174 & 0.098 \\
  L$_{2}$ & 122.9 & 0.243 & 0.098 \\
  L$_{3}$ & 133.0 & 0.248 & 0.097 \\
  L$_{4}$ & 333.5 & 0.171 & 0.054 \\
  L$_{5}$ & 356.2 & 0.125 & 0.026 \\
  L$_{6}$ & 360.6 & 0.337 & 1.901 \\
  L$_{7}$ & 373.1 & 0.069 & 0.430 \\
  L$_{8}$ & 376.4 & 0.076 & 0.783 \\
  L$_{9}$ & 386.7 & 0.074 &  -    \\
  L$_{10}$ & 400.1 & 0.081 & 0.026 \\
\hline\hline
\end{tabular}
\end{table}

Base on our results, the intrinsic large $PF$ is the most important feature for the bulk PdS. Since only nominally undoped sample was studied in this initial report, it's not hard to infer that further improvement of the $PF$ should be a good method through optimizing doping, which could change the density of states at the Fermi level depending on the relation\cite{Heremans}: $S$ = $\frac{\pi^2}{3}\frac{k_{B}^2T}{q}[\frac{dln\sigma(E)}{dE}]_{E = E_F}$ ($q$ is the carrier charge, $E_F$ is the Fermi energy). Another way to improve the power factor of PdS is through band structure engineering, as many efforts have been explored in recent years. Such as the $zT$ of PbTe reached 1.5 at 773 K by distortion of the electronic density of states\cite{Heremans}, and a higher $zT$ value of $\sim$1.8 was observed in PbTe$_{1-x}$Se$_x$ alloys by producing the convergence of many valleys at the desired temperatures. Furthermore, a successful approach of rationally tuning crystal structures in non-cubic materials has been proposed, which has enhanced $zT$ values in a few carefully selected chalcopyrites\cite{Hliu}. In addition, it can be expected that the alloying between Pd and S will lead to a reduction in $\kappa_{lat}$ due to the alloy scattering effect.

\begin{figure}[tbp]
\includegraphics[width=\columnwidth]{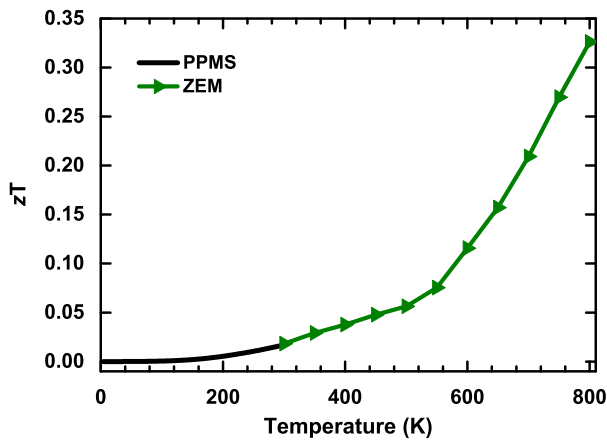}
\caption{Temperature dependence of the dimensionless figure of merit ($zT$) from 2 K to 800 K, The maximum value of $zT$ is 0.33 at 800 K.}
\end{figure}
At the same time, the $\kappa$ of bulk PdS is too large compared with other thermoelectric materials with low thermal conductivity (Fig. 4). Therefore, the reduction of $\kappa$ appears to be very important and useful with the purpose of improving $zT$. Alloying and nanostructuring are both optional routes to increase phonon scatting with the purpose of reducing $\kappa$ effectively\cite{snyder}. Many thermoelectric materials have been improved by these methods, such as the analogous binary PbS, the $zT$ value of the nanostructured PbS is about twice as high as previously reported ($zT \sim 0.4$)\cite{johnsen1}. SiGe is a well-known alloy for high-temperature thermoelectric applications. Recently, by controlling its nanoscale structure, the $zT$ values of both $p$- and $n$-type SiGe have been enhanced\cite{WWang,Joshi}. More importantly, nanostructuring has proven an efficient method to lower thermal conductivities and will not change the electrical transport properties too much\cite{biswas}. Therefore, controlling the nanoscale of binary PdS will be an useful way to improve $zT$ value.

\section{Conclusions}

In summary, measurements of the electrical and thermal transport properties have been performed on the binary PdS in the temperature range of 2 and 800 K. The large value of power factor (27 $\mu$Wcm$^{-1}$K$^{-2}$) and the maximum value of $zT$ with 0.33 at about 800 K indicate the great potential thermoelectric performance as an undoped thermoelectric material. The observed phonon softening offers a natural understanding for the reduction of the lattice thermal conductivity with temperature due to the enhanced phonon anharmonicity. This study also predicts that the $\kappa_{lat}$ could be reduced through increasing phonon scattering which can be realized by nanostructuring, alloying, or doping. These results suggest that the binary bulk PdS has suitable properties as a potential base thermoelectrical material and many PdS based material are expected to have good performance in thermoelectric applications.

\begin{acknowledgments}
Lei Su acknowledged the support from the Natural Science Foundation of China (No. 21273206). Xun Shi and Li-Dong Chen acknowledged the support from the National Basic Research Program of China (973-program) under Project No. 2013CB632501, the Natural Science Foundation of China under the No. 11234012, and the Shanghai Government (Grant No.15JC1400301).
\end{acknowledgments}


\begin{references}

\bibitem{bell} L. E. Bell, Cooling, heating, generating power, and recovering waste heat with thermoelectric systems, Science \textbf{321}, 1457 (2008).

\bibitem{snyder} G. J. Snyder, and E. S. Toberer, Complex thermoelectric materials, Nat. Mater. \textbf{7}, 105 (2008).

\bibitem{sales} B. C. Sales, D. Mandrus, and R. K. Williams, Filled skutterudite antimonides: A new class of thermoelectric materials, Science \textbf{272}, 1325 (1996).

\bibitem{biswas} K. Biswas, J. Q. He, I. D. Blum, C. I. Wu, T. P. Hogan, D. N. Seidman, V. P. Dravid, and M. G. Kanatzidis, High-performance bulk thermoelectrics with all-scale hierarchical architectures, Nature \textbf{489}, 414 (2012).

\bibitem{ypei} Y. Z. Pei, X. Shi, A. LaLonde, H. Wang, L. D. Chen, and G. J. Snyder, Convergence of electronic bands for high performance bulk thermoelectrics, Nature \textbf{473}, 66 (2011).

\bibitem{luxu} X. Lu, D. T. Morelli, Y. Xia, F. Zhou, V. Ozolins, H. Chi, X. Y. Zhou, and C. Uher, High performance thermoelectricity in earth-abundant compounds based on natural mineral tetrahedrites, Adv. Energy Mater. \textbf{3}, 342 (2013).

\bibitem{wancl} C. L. Wan, Y. F. Wang, N. Wang, W. Norimatsu, M. Kusunoki, and K. Koumoto, Development of novel thermoelectric materials by reduction of lattice thermal conductivity, Sci. Technol. Adv. Mater. \textbf{11}, 044306 (2010).

\bibitem{zhaold2} L. D. Zhao, S. H. Lo, J. He, H. Li, K. Biswas, J. Androulakis, C. I. Wu, T. P. Hogan, D. Y. Chung, V. P. Dravid, M. G. Kanatzidis, High performance thermoelectrics from earth-abundant materials: Enhanced figure of merit in PbS by second phase nanostructures, J. Am. Chem. Soc. \textbf{133}, 20476 (2011).

\bibitem{johnsen1} S. Johnsen, J. Q. He, J. Androulakis, V. P. Dravid, I. Todorov, D. Y. Chung, M. G. Kanatzidis, Nanostructures boost the thermoelectric performance of PbS, J. Am. Chem. Soc. \textbf{133}, 3460 (2011).

\bibitem{zhaold} L. D. Zhao, J. He, S. Hao, C. L. Wu, T. P. Hogan, C. Wolverton, V. P. Dravid, and M. G. Kanatzidis, Raising the thermoelectric performance of $p$-Type PbS with endotaxial nanostructuring and valence-band offset engineering using CdS and ZnS, J. Am. Chem. Soc. \textbf{134}, 16327 (2012).

\bibitem{qingt} Q. Tan, L. D. Zhao, J. F. Li, C. F. Wu, T. R. Wei, Z. B. Xing, and M. G. Kanatzidis, Thermoelectrics with earth abundant elements: low thermal conductivity and high thermopower in doped SnS,
J. Mater. Chem. A \textbf{2}, 17302 (2014).

\bibitem{heying} Y. He, T. Day, T. Zhang, H. L. Liu, X. Shi, L. D. Chen, and G. J. Snyder, High thermoelectric performance in non-toxic earth-abundant copper sulfide, Adv. Mater. \textbf{26}, 3974 (2014).

\bibitem{Folmer} J. C. W. Folmer, J. A. Turner, and B. A. Parkinson, Photoelectrochemical characterization of several semiconducting compounds of palladium with sulfur and/or phosphorus, J. Solid State Chem. \textbf{68}, 28 (1987).

\bibitem{Ferrer} I. J. Ferrer, P. D. Chao, A. Pascual, and C. S¨¢nchez, An investigation on palladium sulphide (PdS) thin films as a photovoltaic material, Thin Solid Films \textbf{515}, 5783 (2007).

\bibitem{Barawi} M. Barawi, I. J. Ferrer, J. R. Ares, and C. Sscnchez, Hydrogen evolution using palladium sulfide (PdS) nanocorals as photoanodes in aqueous solution, Acs. Appl. Mater. Inter. \textbf{6}, 20544 (2014).

\bibitem{Bladon} J. J. Bladon, A. Lamola, F. W. Lytle, W. Sonnenberg, J. N. Robinson, and G. Philipose, A palladium sulfide catalyst for electrolytic plating, J. Electrochem. Soc. \textbf{143}, 1206 (1996).

\bibitem{Zubkov} A. Zubkov, T. Fujino, N. Sato, and K. Yamada, Enthalpies of formation of the palladium sulphides, J. Chem. Thermodyn. \textbf{30}, 571 (1998).

\bibitem{yang1} C. H. Yang, Y. Y. Wang, C. C. Wan, and C. J. Chen, A search for the mechanism of direct copper plating via bridging ligands, J. Electrochem. Soc. \textbf{143}, 3521 (1996).

\bibitem{Pascual} A. Pascual, J. R. Ares, I. J. Ferrer, and C. R. Sanchez, {\it International Conference on - ICT}. 376 (2003).

\bibitem{YPei} Y. Pei, J. Lensch-Falk, E. S. Toberer, D. L. Medlin, and G. J. Snyder, High thermoelectric performance in PbTe due to large nanoscale Ag$_2$Te precipitates and La doping, Adv. Funct. Mater. \textbf{21}, 241 (2011).

\bibitem{Ravich} Y. I. Ravich, B. A. Efimova, and I. A. Smirnov, {\it Semiconducting Lead Chalcogenides} (Plenum, New York, 1970) (and references therein).

\bibitem{Scanlon} W. W. Scanlon, in Solid State Physics (Edited by F Seitz and D Turnbull), Vol 9, p 83. ({\it Academic, New York}) (1959).

\bibitem{Kumar} G. S. Kumar, G. Prasad, and R. O. Pohl, Experimental determinations of the Lorenz number, J. Mater. Sci. \textbf{28}, 4261 (1993).

\bibitem{Gofryk} K. Gofryk, D. Kaczorowski, T. Plackowski, J. Mucha, A. Leithe-Jasper, W. Schnelle, and Y. Grin, Magnetic, transport, and thermal properties of the half-Heusler compounds ErPdSb and YPdSb, Phys. Rev. B \textbf{75}, 224426 (2007).

\bibitem{Dilley} N. R. Dilley, E. D. Bauer, M. B. Maple, S. Dordevic, D. N. Basov, F. Freibert, T. W. Darling, A. Migliori, B. C. Chakoumakos, and B. C. Sales, Thermoelectric and optical properties of the filled skutterudite YbFe$_4$Sb$_{12}$ Phys. Rev. B \textbf{61}, 4608 (2000).

\bibitem{cwli} C. W. Li, J. Hong, A. F. May, D. Bansal, S. Chi, T. Hong, G. Ehlers, and O. Delaire, Orbitally driven giant phonon anharmonicity in SnSe, Nat. Phys. \textbf{11}, 1063 (2015).

\bibitem{DHwang} D. H. Wang, T. H. Xu, and H. Y. Song, Thermal expansion behaviors of epitaxial film for wurtzite GaN studied by using temperature-dependent Raman scattering, Acta Phys. Sin. \textbf{65}, 130702 (2016).

\bibitem{dongz} L. D. Zhao, S. H. Lo, Y. S. Zhang, H. Sun, G. J. Tan, C. Uher, C. Wolverton, V. P. Dravid, and M. G. Kanatzidis, Ultralow thermal conductivity and high thermoelectric figure of merit in SnSe crystals, Nature \textbf{508}, 373 (2014).

\bibitem{Nielsen} M. D. Nielsen, V. Ozolins, and J. P. Heremans, Lone pair electrons minimize lattice thermal conductivity, Energ. Environ. Sci. \textbf{6}, 570 (2013).

\bibitem{zhge} Z. H. Ge, B. P. Zhang, and Z. X. Yu, Effect of spark plasma sintering temperature on thermoelectric properties of Bi$_2$S$_3$ polycrystal, J. Mater. Res. 26, 2711 (2011).

\bibitem{Heremans} J. P. Heremans, V. Jovovic, E. S. Toberer, A. Saramat, K. Kurosaki, A. Charoenphakdee, S. Yamanaka, and G. J. Snyder, Enhancement of thermoelectric efficiency in PbTe by distortion of the electronic density of states, Science \textbf{321}, 554 (2008).

\bibitem{Hliu} J. W. Zhang, R. H. Liu, N. Cheng, Y. B. Zhang, J. H. Yang, C. Uher, X. Shi, L. D. Chen, and W. Q. Zhang, High-performance pseudocubic thermoelectric materials from non-cubic chalcopyrite compounds, Adv. Mater. \texttt{26}, 3848 (2014).

\bibitem{WWang} X. W. Wang, H. Lee, Y. C. Lan, G. H. Zhu, G. Joshi, D. Z. Wang, J. Yang, A. J. Muto, M. Y. Tang, J. Klatsky, S. Song, M. S. Dresselhaus, G. Chen, and Z. F. Ren, Enhanced thermoelectric figure of merit in nanostructured $n$-type silicon germanium bulk alloy, Appl. Phys. Lett. \textbf{93}, 193121 (2008).

\bibitem{Joshi} G. Joshi, H. Lee, Y. Lan, X. Wang, G. Zhu, D. Wang, R. W. Gould, D. C. Cuff, M. Y. Tang, M. S. Dresselhaus, G. Chen, and Z. F. Ren, Enhanced thermoelectric figure-of-merit in nanostructured $p$-type silicon germanium bulk alloys, Nano Lett. \textbf{8}, 4670 (2008).

\end{references}
\end{document}